# Publish/Subscribe-enabled Software Defined Networking for Efficient and Scalable IoT Communications

Akram Hakiri, Pascal Berthou and Aniruddha Gokhale

*Abstract*— The Internet of Things (IoT) is the result of many different enabling technologies such as embedded systems, wireless sensor networks, cloud computing, big-data, etc. used to gather, process, infer, and transmit data. Integrating all these technologies requires a comprehensive and holistic research effort to address all the challenges imposed by these technologies, especially for sensing and delivering information from physical world to cloud-hosted services. In this paper, we outline the most important issues related to standardization efforts, mobility of objects, networking and gateway access, and QoS support. In particular, we describe a novel IoT network architecture that integrates Software Defined Networking (SDN) and the Object Management Group's Data Distribution Service (DDS) middleware. The proposed architecture will improve service delivery of IoT system and will bring flexibility to the network.

## I. INTRODUCTION

The widespread evolution of the Internet of Things (IoT) concept imposes complex requirements on both the underlying networks and communication mechanisms between heterogeneous smart objects that communicate over the Internet. The current estimate for the number of deployed things are on the order of 50 billion connected devices and by 2020 will encompass 1000 times more connected mobile devices, all with different requirements, which will have a big impact on how people will interact with the surrounding things. This is however a very difficult goal to achieve because today's network is limited in its ability to address the requirements of even current IoT deployments. In the future, leveraging the IoT potential will be the key enabler to the creation of different applications areas, including smart cities, e-health, industrial, transportation, retail, safety, and environmental services.

In order to achieve this goal, it is necessary to provide an efficient network architecture that aids in developing many scalable, interoperable and predictable IoT applications. This network architecture must be flexible enough to be reprogrammed in accordance with any change in IoT application needs. IoT usually adopts different communication schemes compared to the traditional Internet stemming primarily from the variety of resource-constrained hardware platforms. In particular, IoT network is a highly unstructured cloud of wireless and potentially mobile devices, whose states change dynamically (e.g., sleeping and waking up, connected and/or disconnected) as well as their locations and speeds change. These smart applications should run in resource-constrained devices with limited computation resources, small data-storage capabilities and low-power consumption. Since these devices sense the environment and communicate, they should be autonomic to respond to different scenarios without human interaction.

Since these large number of smart objects are being provided as services connected to the Internet, the current IPv4 addressing no longer suffices since it has reached its limit and hence efforts are underway in deploying IPv6 addresses. Additionally, the heterogeneous and dynamic aspects of IoT systems pose major challenges for the underlying network by requiring support for handling heterogeneity, dynamic changes, device discovery as well as context-awareness. Contemporary network protocols are designed in isolation to solve a specific problem and are often retrofitted to address a new requirement. Unfortunately, this approach has limitations for IoT. Moreover, existing protocols often lack the right abstractions that address the requirements of IoT communication. Thus, if network resource utilization is a concern, the network must be flexible enough to be reprogrammed in accordance with any change in IoT application needs. Current network provisioning approaches neither address the dynamicity of IoT applications nor care about resource utilization. Recently, Software Defined Networking (SDN) [1] was introduced to deliver dramatic improvements in the network agility and flexibility. The SDN paradigm is thus a promising solution to solve the resource management needs of the IoT environment, however, it cannot address the heterogeneous and dynamic needs of IoT applications.

At the distributed systems-level, the Web of things (WoT) [2] concept was introduced to alleviate the heterogeneity issues by allowing smart devices to speak the same language based on open web technologies such as HTTP and REST principles for information sharing. However, to fully explore the potential of WoT, many challenges remain unresolved such as security and scalability. Moreover, WoT cannot address the need for distributed, peer-to-peer, publish/subscribe semantics, which are a key requirement for IoT applications.

Many solutions at the middleware level were introduced for IoT applications. LinkSmart [3] supports resource discovery, description and access based on XML and web protocols. OpenIoT [4] realizes on-demand access to cloud-based IoT services through Internet-connected smart objects. These contributions focus on the upper layer problems such as enabling IP-based radio communication over middleware, however, without paying attention to the underlying network. The Object Management Group's Data Distribution Service (DDS) provides real-time, scalable, data-centric publish/subscribe capabilities. However, although DDS has been used to develop many scalable, efficient and predictable applications at a local area network-scale, its QoS mechanisms are rarely propagated to the network layer, which impedes its use across wide area scale. Moreover, DDS is not a network-level solution.

To address both the network- and distributed systems-level concerns in IoT, we propose combining ideas from SDN with the message-oriented publish/subscribe DDS middleware to define a powerful and simple abstract layer that is independent of the specific networking protocols and technology.

The remainder of this paper is organized as follows: Section II briefly introduces the concepts of SDN and DDS. Section III introduces the architecture of IoT and provides a glimpse into the most important open issues related to the deployment of IoT networks. Section IV describes the architecture of our SDN solution for efficient use of IoT and discusses the role of the proposed solution in solving the above-mentioned issues. Finally, Section V provides concluding remarks describing potential future directions and open research problems in this realm.

## II. BACKGROUND

### A. Software Defined Networking (SDN) and Network Function Virtualization (NFV)

Software-Defined Networking (SDN) has emerged as a new intelligent architecture for network programmability. It moves the control plane outside the switches to enable external centralized control of data through a logical software entity called controller. The controller offers northbound interfaces to network applications that provide higher level abstractions to program various network-level services and applications. It also uses southbound interfaces to communicate with network devices. OpenFlow is an example of southbound protocols. OpenFlow's behavior is simple but it can allow complex configurations: the hardware processing pipeline from legacy switches is replaced by a software pipeline based on flow tables. These flow tables are composed of simple rules to process packets, forward them to another table and finally send them to an output queue or port.

One complementary technology to SDN called Network Function Virtualization (NFV) has the potential to dramatically impact future networking by providing techniques to refactor the architecture of legacy networks by virtualizing as many network functions as possible. NFV advocates the virtualization of network functions as software modules running on standardized IT infrastructure (like commercial off-the-shelf servers), which can be assembled and/or chained to create services.

### B. OMG Data Distribution Service (DDS)

The Object Management Group has standardized the Data Distribution Service (DDS) [5] middleware as a protocol for the IoT to enable network interoperability between connected machines, enterprise systems, and mobile devices. DDS can be deployed in platforms ranging from low-footprint devices to the cloud and supports efficient bandwidth usage as well as agile orchestration of system components. DDS provides a flexible and modular structure by decoupling: (1) location, via anonymous publish-subscribe, (2) redundancy, by allowing any numbers of readers and writers, (3) time, by providing asynchronous, time-independent data distribution, (4) message flow, by providing message-based data-centric connection management, and (5) platform, by supporting a platform-independent model that can be mapped to different platform-specific models, such as C++ running on VxWorks or Java running on Real-Time Linux.

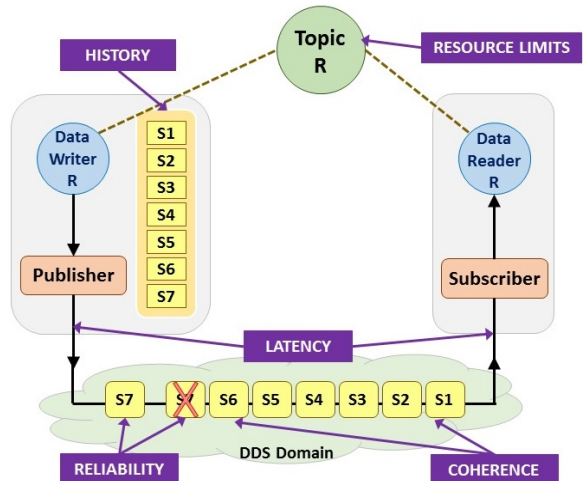

Figure 1: A View of the Data Distribution Service

Figure 1 depicts the relation between DDS entities: domains, topics, publishers, data writers, subscribers, and data readers. A DDS domain represents a virtual global data-space; information provided in the domain are accessible by the applications registered to that domain. Publishers manage one or more data writers and subscribers manage one or more data readers. Also, DDS recognizes the importance of discovery and of meta-data, two areas addressed by IoT systems. The OMG DDS standard has evolved over time. Initially, it provided only the platform- and language-independent mechanisms to build distributed publish/subscribe systems with QoS capabilities. Subsequently, to promote interoperability, the standard provided a discovery mechanism via the Real-Time Publish-Subscribe (RTPS) protocol [6]. During the discovery process each domain participant maintains a local database about all the active DataWriters and DataReaders that are in the same DDS domain. Several additional enhancements to the standard are being discussed including the extended and dynamic types for topics, remote method invocation, integration with Web protocols, security extensions, and integration with SDN at the northbound APIs.

### III. NETWORK COMMUNICATION CHALLENGES IN IoT SYSTEMS

In this section we first describe the high level architecture of IoT systems. Then, we describe their most challenging networking open issues.

*A. IoT Architecture*

Figure 2 shows a high-level architecture of IoT systems, which is composed of three domains: the Device Domain, the Network Domain and the Application Domain. In the device domain, the device provides direct connectivity to the network domain via access networks, which may include limited range PAN technologies such as Bluetooth, ZigBee, etc., or via a gateway that acts as a network proxy for the network domain.

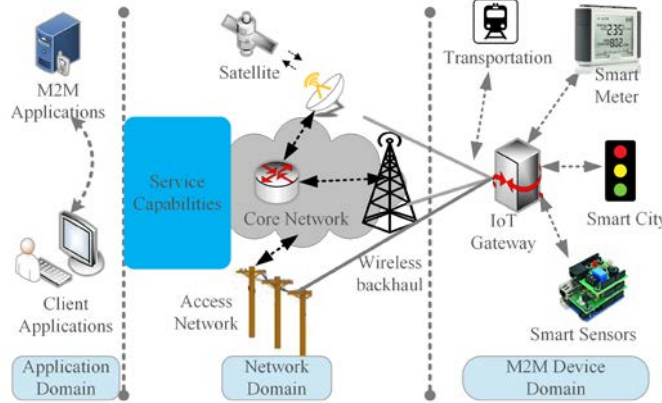

Figure 2: High-level IoT architecture

Such a gateway must be flexible enough to efficiently manage available resources, QoS, security, as well as multimedia data exchange. These gateway concepts are prevalent in home ADSL models and WiFi access points found in cyber cafes and wireless hotspots. Because IoT systems integrate heterogeneous smart objects, the design of the gateway is quite different because it must not require each IoT subnetwork to have its own gateway. Thus, a convergent architecture towards a unique solution that integrates traffic incoming from heterogeneous smart devices should be designed. Furthermore, as smart objects are resource- and energy-constrained, the gateway should be aware of the context of each process being managed. It should also employ intelligent routing protocols and caching techniques to route the traffic across the less constrained paths.

The network domain includes different access networks, which provide connectivity through diverse technologies, such as xSDL, Satellite, etc. to devices and/or gateways. It also provides connectivity to the core network that includes heterogeneous and multi-technology connectivity, such as 3GPP, TISPAN, and LTE-A. Finally, the application domain includes the IoT applications and server/cloud infrastructures. The latter have to share their content, possibly back them up to other devices, analytic programs, and/or people who need to monitor real-time response to events. They also include service capabilities, which provide functions shared between different applications through open, high-level abstractions and interfaces that hide the specificities of the underlying networks.

*B. Network-level challenges for IoT*

We now describe five key network-level challenges, active research areas, and standardization efforts for IoT

*1) Current Standardization Efforts*

Several initiatives around the world taken by academic organizations, industries, standardization bodies, and governments have emerged during the past few years to enable the IoT deployment in everyday life. Among them, the 6LowPAN protocol was introduced to reside between the IPv6 and MAC protocol layers to make IPv6 protocol compatible with low capacity devices. The Routing Over Low-Power and Lossy Networks (ROLL) effort focuses on the routing issues for lower-energy consumption smart object networks. The Constrained Application Protocol (CoAP) is a specialized web transfer protocol for constrained smart objects and networks. Similarly, Machine-to-Machine (M2M) promotes the development and the maintenance of an end-to-end architecture for M2M, including sensor network integration, naming, addressing, location, QoS, security, etc. These efforts remain isolated, however.

*2) Mobility Management*

A large number of IoT devices are not fixed but manifest in diverse mobile scenarios. Therefore, supporting and managing efficient mobility for those smart objects is of primary interest to the IoT community. Mobility management in IPv6 networks may be based on either a home agent such as in Mobility IP or home location registrar/visitor location registers. The former does not require the use of central server, which is critical in terms of scalability and flexibility. The latter is widely used in cellular networks, but is not suited for IoT communication since it requires additional and dedicated infrastructure to manage mobility. Both approaches generate a huge amount of signaling traffic to discover devices and maintain their up to date positions/locations, which degrades the network performance. Given the frequent mobility of IoT devices, it is challenging for a SDN controller to have a network view about the mobility of IoT objects to manage their spatial-temporal requests, collaborate with other controllers for adaptive handover and dynamic flow scheduling in IoT networks

*3) Recurring Distributed Systems Issues*

Middleware is required in the IoT environment to provide reusable solutions to frequently encountered problems like heterogeneity, interoperability, security, and dependability. The Message Queuing Telemetry Transport (MQTT) [7] was introduced as a reliable and lightweight messaging protocol for low-bandwidth, and high-latency smart devices. MQTT enables device-to-device communication through a centralized broker. However, the centralized broker presents a single point of failure. Moreover, maintaining a TCP connection between the client and the broker at all times becomes problematic for environments where packet loss is high and computation resources are scarce.

Of particular interest is the pub/sub data-centric middleware because it offers common services to help deliver events from source nodes to interested destinations in an asynchronous way. This kind of middleware integrates diverse types of communication patterns expected for IoT systems. The key challenge is how to cope with congestion control that arises in many M2M scenarios. A novel solution should be investigated to take into account the requirements of IoT devices such as short packet length and short-lived.

*4) Communication Protocols*

The major goals of widely used transport protocols such as TCP are to guarantee end-to-end reliability as well as congestion control. However, TCP is unsuitable for IoT scenarios since it incurs substantial overhead during the connection setup and several performance degradation over wireless channels during congestion control. It also requires data buffering at both sources and sinks for packet retransmission and/or in-order delivery. Conversely, UDP has been used for CoAP due to its low overhead and connectionless properties. With gateways acting as repeaters and/or protocol translators, devices are expected to contribute to QoS management by optimizing the resources utilization of IoT networks. Thus, significant research efforts must be devoted to address the particular issues related to reliable transport, routing and QoS provisioning that arises in many IoT scenarios.

*5) Security and Privacy*

The current landscape of IoT uses diverse wireless and/or cellular communication technologies which makes eavesdropping and vulnerability of channels extremely simple. Also, IoT devices have limited power and computation resources making complex security schemes infeasible. In particular, various properties, such as confidentiality, integrity, and privacy must be ensured. Security policies and mechanisms should be specified to guarantee privacy. They should be configurable to suit individual ways to control which of their personal data is being collected, who is collecting those data, and which operations will be performed on such a data. However, in IoT there is no such infrastructure or servers to manage authentication that achieve the appropriate security policies. Thus, none of the existing solutions that have been proposed for wireless sensor networks can be applied to IoT.

## IV. SDN-DDS Architecture for IoT

This work considers the typical architecture of IoT system described in Figure 2, where sensors and actuators are connected to local processing and to the Internet. The Internet access is provided through a core router connected to service provider through terrestrial or mobile access (i.e., 3G/4G, ADSL, etc.). The router communicates with IoT gateways to enable connectivity over a wide range of communication technologies to support Wi-Fi, Ethernet, VPN, USB, Serial, ZigBee, etc. Both smart objects and gateways use DDS middleware to publish/subscribe data. To allow IoT system supporting SDN, we added DDS northbound interface to the SDN controller that exposes all the necessary functionalities of OpenFlow to the controller to build network-agnostic support for IoT systems.

*A. The Controller Architecture*

The controller integrates a DDS messaging layer that in turn acts as a mediator between the IoT system and the network. It supports both the proactive and reactive flow programming without the need for complex RESTful interfaces nor OSGI services. In contrast to RESTful interfaces that use synchronous client-server point-to-point communication, which may hinder scalability, DDS pub/sub messaging allows anonymous, asynchronous and many-to-many communication semantics.

As depicted in Figure 3, three services are implemented by the mediation layer: the *Packet Handler* uses the DDS DataReader (DR) Listener to receive the PACKET_IN events captured by the controller after being forwarded by the SDN data plane. The *Packet Forwarder* service enables forwarding packets received through PACKET_IN events or new packets created by IoT applications. The latter are encapsulated in DDS Topics, then sent to the *Packet Forwarder* service through the publisher and its DataWriter (DW) Listener. Finally, the *Flow Programming* service is used to define the flow programming rules on the OpenFlow switches.

The mediation layer transports messages between the network control application and SDN services in the controller. That is, the DDS middleware natively fulfills the requirements of the future proactive IoT in SDN communication, which makes it an attractive technology to satisfy the features they need, such as scalability, reliability, flexibility, security and real-time data delivery.

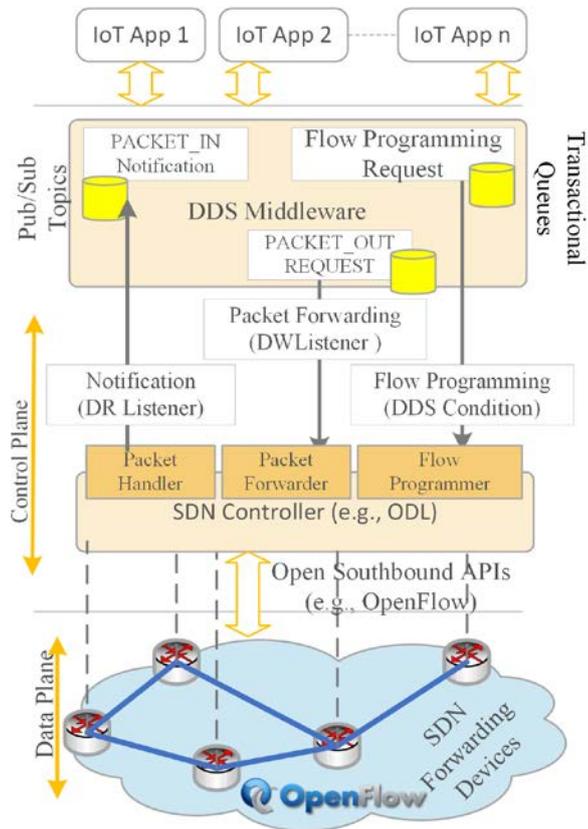

Figure 3: A view of the SDN controller with the DDS layer

B. *Addressing Open Research Challenges*

The proposed architecture addresses the research questions discussed earlier as follows:

  1) *Standardization and Open Innovation*

The proposed architecture is based on SDN and DDS, which are open and standardized that will minimize or even prevent the lock-in effect. As the middleware choice in our architecture, DDS can interoperate with existing standardized protocols such CoAP, which can be easily plugged in as an extensions to the middleware without adding any complexity to the underlying network. Moreover, network virtualization enabled SDN as exemplified by OpenFlow will allow experimenting with new ideas. Using a modular design with some inexpensive general hardware platform and customized switch control software, it will be possible to improve the performance of the network.

  2) *Addressing and Mobility*

Both DDS and SDN help enable a programmable wireless data plane to allow mobility management, dynamic channel configuration, and rapid client association. First, DDS enables many-to-many, broker-less IoT communication so that smart objects can benefit from self-addressable and self-routable data. It also enables bounded use of resources over potentially intermittent links, and makes it possible to replay messages upon reconnection. Second, SDN techniques are being applied to wireless sensor networks to improve resource and mobility management [8].

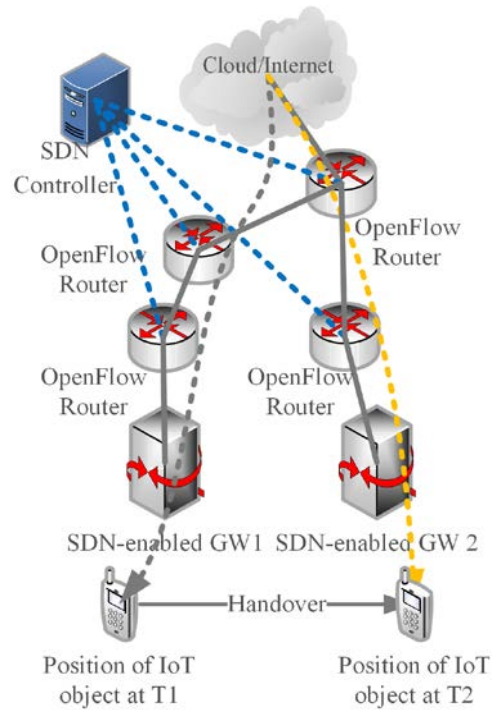

Figure 4: Use of SDN controller to improve mobility management

We propose to virtualize IoT gateways to maintain active session continuity for IoT devices. The gateway virtualization helps in improving IoT multi-homing by enabling dynamic attachment of sensors to multiple networks. Figure 4 illustrates an example of handover migration of mobile smart devices from SDN-enabled gateway (GW1) to another gateway (GW2) in a wide area network. If at any time, a smart devices moves between different networks, then the centralized SDN controller modifies flow-tables in OpenFlow switches to take into account the mobility of the device.

*3) Efficient Middleware Integration*

Using a single middleware that supports multiple communication patterns is a very cost-effective way of developing and maintaining large, distributed IoT systems. DDS provides message-oriented publish/subscribe semantics that implement various communication patterns including transactional queues for request/response interaction (proactive flow programming), guaranteed delivery, and publish/subscribe topics for event-based interaction (reactive flow programming). Thus, we can program flows through a request-response interface implemented by message queues and react to packet-in events by subscribing to events through message topics. Moreover, DDS supports traffic differentiation and prioritization along with selective dawn-sampling. It can mark DSCP field of IP packets sent by smart sensors to enable end-to-end QoS provisioning.

*4) Service Discovery and Interoperability*

DDS discovery helps different smart devices to find each other. The IoT gateway maintains a string formatted local objects list to build a complete picture about all its surrounding devices. The list includes the location and the description of each IoT object in the network, e.g., IP addresses, CoAP URI, and multicast address. However, in some cases multicast is not available in IoT networks. So, if discovery frames sent with multicast addresses cannot be handled by the gateway, they will automatically be redirected to the controller. The later will install a correct rule in the gateway to simulate a broadcast network so the next frame will be handled without being sent to the controller. This concept is extended to improve the interoperability of incompatible protocols. For instance, when a non DDS compliant IoT gateway is connected to the wireless AP, the gateway frames are redirected to the controller that also redirect it to an SDN application that encapsulates data into DDS frames according to a predefined scheme.

*5) Scalability handling*

A usual way to connect IoT devices consists of bridging devices with a gateway that collects data. This simple architecture, although frequently used, is not scalable when multiple gateways are used. The main problem stems from the bottleneck when several nodes have to send data at the same time. Usually, data filtering and data fusion are used to cope with this problem. Data filtering aims at dropping unnecessary packets according to a given policy. Data fusion consists of combining data to reduce the number of forwarded packets while keeping almost the same information quality.

DDS offers Fusion and Filtering mechanisms to reduce useless traffic. The former, called batching mode, enables collecting multiple small data samples and sending them in a single network packet to increase the bandwidth effectiveness. The latter is called Content Filtered Topic which makes it possible for a smart device to subscribe to a given Topic and at the same time specify that it is interested in a subset of this topic data. For example, suppose a Topic that contains temperature sensor samples are published with values from 0° to 100°, but the subscribing device needs only values that exceed some threshold. The Content Filtered Topic mode can be used to limit the number of data samples the subscriber has to process and reduce the amount of data sent by the network.

As shown in Figure 5, OpenFlow completes these mechanisms with an efficient monitoring system. The controller can collect traffic statistics at any granularity and is configured with flow entries downloaded to the switches. Byte or packet counters are associated with every OpenFlow entry. It is then easy to get informed of a network flooding and react accordingly.

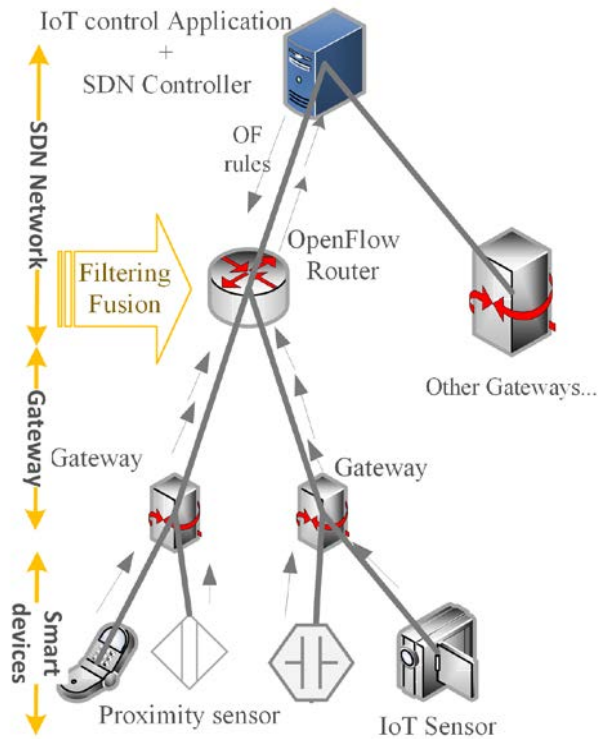

Figure 5: Scalability Handling

.
*6) Context aware network management*

IoT and especially embedded smart sensors tend to generate huge volumes of data that must be transmitted over the network. It is not always easy to configure the transmission scheme to select exactly the relevant sensor data. DDS offers a useful mechanism called multi-channel.

Multichannel Data Writers allows reducing network traffic by subdividing a data flow into a set of streams to be sent over multiple channels. Each channel maintains a set of multicast addresses, defined by application-specified predicate, called filter expression.  It is then possible to associate different multicast addresses with filter expressions, so that only the Topics data that match the expression are delivered to the subscriber.

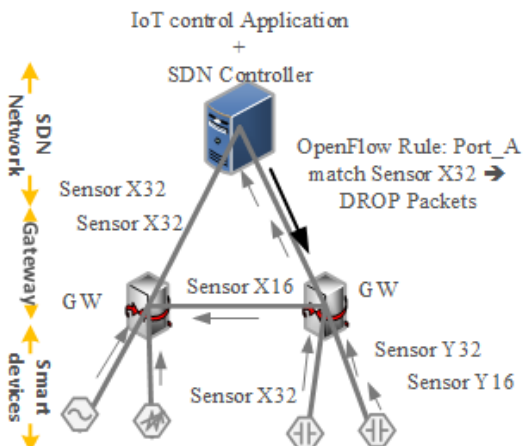

Figure 6: Multi-channel and data filtering

However, multichannel should be configured inside each sending device and cannot be modified on-the-fly easily. OpenFlow could be used to offer a similar service. In Figure 6, the OpenFlow match-action (Openflow rule) treatment chain could be set to filter data according a keyword contained inside the sent data as much as possible closed to the source. The decision can be taken at the highest application level: the application inside the controller. Owing to the OpenFlow monitoring system, the relevant information can be set.

*7) Solving Security issues*

Security issues can be handled by both DDS and SDN. At the middleware level, smart devices can leverage the DDS Security Model (SM) [9], which offers simple and interoperable security policies without compromising the flexibility, scalability, performance, and QoS-awareness offered by DDS. Thus, we can build a fine-grained secure system that grants permissions to DDS domains, Topic or even data object instances within the Topic. Also, DDS "partition" QoS provides another way to create isolated subdomains to defend IoT networks against attacks. Similarly, SDN enables slicing the IoT infrastructure into multiple virtual partition, so that if any attack occurs in a given partition, other IoT partitions remain isolated and secure. Along with network slicing, the controller can program the IoT gateways to conduct fine-grained packet inspection on traffic passing through IoT devices [10]. These statistics collected periodically offer a real-time view of the network state. If any vulnerability occurs in the IoT system, a security SDN application can be deployed on-the-fly at the controller for detecting and driving mitigation of malware and DDoS attacks.

V. CONCLUSION

The Internet of Things (IoT) promises to have a big impact by adding a new dimension in the way people will interact with the surrounding things, and to form a virtual continuum of interconnected smart objects in a worldwide dynamic network. In this paper, we have surveyed the most important challenges that need to be tackled for efficient support of IoT systems. Then, we introduced a data-centric architecture based on a symbiotic relationship between DDS and SDN that enables agile and flexible network orchestration.

As IoT covers a huge range of industries and scales of applications, IoT solutions and architectures are undergoing evolution, with convergence of technologies, and new innovative solutions, such as SDN, will be considered particularly with 5G networks on the horizon (2020 and beyond). As part of our ongoing work for DDS/SDN architecture, we believe that this study can shed light on how we can integrate DDS middleware and SDN to improve IoT communication and promote the adoption of SDN for the future IoT networks.